# FLUCTUATION EFFECTS IN METASTABLE STATES NEAR FIRST ORDER PHASE TRANSITIONS

Dimo I. Uzunov

Institute of Solid State Physics, Bulgarian Academy of Sciences, BG-1784 Sofia, Bulgaria.

## Abstract

Fluctuation effects at first order phase transitions driven by changes of other-than-temperature factors like pressure, concentration, or external fields are investigated by perturbation theory. The results for the fluctuation contributions to the order parameter, the internal energy and the free energy at pre-transitional states near spinodal points of first order phase transitions are presented to the first-non-vanishing order of the expansion parameters of the theory.

## 1. Introduction

Fluctuations in metastable states of systems near equilibrium and ensuing phase transitions are important in many areas of physics. Quantum fluctuations due to Heisenberg's uncertainty principle could have produced small perturbations in the space-time that shaped our Universe. Phase transitions, occurred in the early Universe as it cooled, could have produced the forces and particles we observe today. On the other hand, *thermal fluctuations* (fluctuations in a system subjected to thermal noise) are important in many practical situations. Although metastable states of matter, e.g., supercooled or superheated water, diamond, etc, survive for a period of time, eventually these states transfer to stable states: ice, steam, graphite, etc. For the period of time while the metastable states survive the decomposition, they may be considered as stable ones and their fluctuations bear all the features of equilibrium fluctuations. Usually, fluctuations around stable thermodynamic states (phases) near phase transitions driven by temperature variations are studies. In contrast, here we investigate fluctuation effects in metastable states around equilibrium points of first order phase transitions at fixed temperature when the phase transition occurs by variations of another ("non-temperature") thermodynamic parameter.

From the stand point of the theory of phase transitions the above mentioned examples belong to the category of first-order transitions driven away from the point of equilibrium. One of the best approaches to consider these processes is the Ginzburg-Landau (GL) theory, which is based on the concept of order parameter as the main characteristic of the phase transition and the phases [1-4]. The generalized GL free energy (fluctuation Hamiltonian) is a functional of the order parameter field and can be suitably chosen to incorporate the most essential thermodynamic features of the system including the relevant fluctuation effects.

Traditionally, fluctuations around stable thermodynamic states have been studied in the framework of GL Hamiltonians of the type 2-4 where the numbers indicate the powers of the order parameter in the truncated power-series expansion [5, 6]. In the fluctuation theory of phase transitions [2, 3] the quadratic term is often called "Gaussian term" whereas any non-quadratic term is called "interaction term", e.g., the quartic term in 2-4 Hamiltonians. The quadratic term describes free (non-interacting) fluctuation modes whereas the terms of higher order in powers of the order parameter field describe the relevant fluctuation interactions in the system [2, 3].

The usual 2-4 theories describe standard second order phase transitions where metastable states do not appear. To study metastable states and related problems of materials physics, e.g. phase nucleation, growth, and coarsening, many researchers use the GL approach to Hamiltonians of type 2-3-4, where an additional term of third order in the order parameter is included into consideration [3,6]. Hamiltonians of this type describe first order phase transitions between stable phases. Under certain thermodynamic conditions the same phases may appear by metastable states or to coexist in equilibrium. Owing to the availability of the cubic interaction term, the fluctuation properties of the 2-3-4 Hamiltonians are quite different from those of the more common 2-4 Hamiltonians. Another essential difference with common cases arises in systems when the phase transition is not driven by the temperature but rather by the variations of another thermodynamic parameter. In such cases, the (Landau) parameters of the 2-3-4 theory have unusual temperature dependence and, hence, some thermodynamic properties, including fluctuation effects will differ from those corresponding to more common cases.

In order to investigate such cases, in this paper we introduce an essentially new dependence of the 2-3-4 Hamiltonian on the temperature and the thermodynamic parameter whose variations transform the thermodynamic states from one another and, hence, are responsible for the phase transition of first order. This model could be used to describe a variety of real systems undergoing a certain type of first order phase transitions and metastable phases which occur at fixed temperature and are driven by other-than-temperature factors (pressure, concentration, etc.). Our attention is focused on the fluctuation properties of the metastable states. We calculate the free energy, the internal energy, and the order parameter outside the close vicinity of the spinodal point where the fluctuations are anomalously strong and the perturbation theory breaks down [3].

The paper is organized as follows. In Sec. 2 we introduce the new GL model and present the results from the mean-field approximation (MFA). The MF results are important for the outline of the phase diagram, the location of the phases, and allow for to distinguish the net fluctuation effects set forth in the paper. The fluctuations are investigated in Gaussian approximation (GA) as well as beyond it by taking into account the fluctuation interactions up to second order of the perturbation theory (Sec. 3). The perturbation theory approach is outlined and the final results are presented in Sec. 3. In Sec. 4 we summarize our findings.

## 2. Ginzburg-Landau Hamiltonian and mean-field results

We consider a typical first order phase transition, e.g., crystallization, driven by variations of thermodynamic parameters such as pressure, concentration, or external fields. These variations will be described by a dimensionless parameter $\Delta$; henceforth referred to as "supersaturation", or, "driving force" of the phase transition. Presence of the thermal noise in the system leads to variations of the scalar order parameter $\eta(\mathbf{x}) = \eta_e + \delta\eta(\mathbf{x})$, where $\eta_e$ is the equilibrium part of the order parameter and $\delta\eta(\mathbf{x})$ is the fluctuation part of the total field (the spatial vector $\mathbf{x} \in V$ runs values in the volume $V$ of the system). In the most common examples of real systems the equilibrium order parameter $\eta_e$ is uniform whereas the fluctuation field $\delta\eta(\mathbf{x})$ contains both uniform and spatially ($\mathbf{x}$-) dependent components and here we focus the attention to this usual case.

The general GL Hamiltonian of such systems with short-range intermolecular interactions can be written in the general form [1-4]

$$\mathcal{H}[\eta(\mathbf{x})] = \int_V d^3x \left\{ H[\eta(\mathbf{x}), \Delta] + \frac{1}{2}\kappa[\boldsymbol{\nabla}\eta(\mathbf{x})]^2 \right\}, \tag{1}$$

where the integration over the spatial vector $\mathbf{x}$ is performed in the volume $V$ of a three dimensional system, the gradient term with the nabla operator $\boldsymbol{\nabla} = (\nabla_1, \nabla_2, \nabla_3)$) and the stiffness parameter $\kappa$ represents the energy of the spatial variations whereas the Hamiltonian density $H[\eta(\mathbf{x}), \Delta]$ represents the free energy density which survives when the spatial fluctuations of the field $\eta(\mathbf{x})$ are ignored, namely, in case of a simple (uniform) field $\eta(\mathbf{x}) \approx \eta$; henceforth the notation $\eta$ stands for $\eta = \eta_e + \delta\eta$, where $\delta\eta$ is the unavoidable uniform fluctuation.

In MFA, $\eta(\mathbf{x}) \approx \eta_e$ [2, 3], the fluctuations are neglected by ignoring both the spatial dependence of the order parameter and the uniform fluctuation $\delta\eta = (\eta - \eta_e)$ contained in the total (non-equilibrium) order parameter $\eta$. Let us remind that in case of uniform order parameter $\eta$, the energy density $H(\eta, \Delta)$ is the free energy per unit volume in the standard Landau theory [3, 6], $F/V = H(\eta, \Delta)$, whereas the Hamiltonian with the total field $\eta(\mathbf{x})$ is often referred to as generalized free energy functional, or, GL functional provided the density $H[\eta(\mathbf{x}), \Delta]$ is expanded in Landau series in powers of $\eta(\mathbf{x})$ and truncated at certain, usually fourth or sixth order, of the field $\eta(\mathbf{x})$ [2, 3]. While $\eta(\mathbf{x})$ denotes any physically allowed configuration of the order parameter field, including the uniform configuration, $\eta(\mathbf{x}) = \eta$, the GL functional presents a nonequilibrium free energy. The equilibrium values of the free energy are given by those configurations of the filed $\eta(\mathbf{x})$ which correspond to extrema of the GL functional, i.e., to solutions of the equation $\delta H\{\eta(\mathbf{x}),\Delta\}/\delta\eta(\mathbf{x}) = 0$.

Within the MF approximation [2, 3], the equilibrium values $\eta_e$ of the total uniform order parameter $\eta$ are obtained as solutions of the equation $\partial H(\eta,\Delta)/\partial\eta = 0$ and the MF free energy is given by $F = VH(\eta_e,\Delta)$. Besides, we remind that the extrema of the GL functional could be of three types: (1) global minimum, namely, the stable phase, (2) relative minima (metastable phase), and (3) maxima or inflection points, which correspond to unstable phases. These stability properties of the phases may change with variations of the theory parameters, in our case, the parameter $\Delta$. For example, suitable variations of the tuning parameter $\Delta$ may transform the stability feature of a given phase from stability to metastability, or, even to instability, and vice versa. Phases which occur in real situations are the stable phases (under usual thermodynamic conditions) and metastable phases (under some particular conditions). The unstable phases are practically impossible to occur and therefore uninteresting.

To obtain concrete results we should shape the density $H(\eta,\Delta)$ of a particular form. A broad set of first order phase transitions with scalar order parameter field can be described by the 2-3-4 Hamiltonian density [1-4]

$$H[\eta(\mathbf{x}),\Delta] = H_\alpha + \frac{1}{2}W\left\{(1-\Delta)\eta^2(\mathbf{x}) - 2\left(1-\frac{\Delta}{3}\right)\eta^3(\mathbf{x}) + \eta^4(\mathbf{x})\right\} \quad (2)$$

Here $H_\alpha$ is the Hamiltonian density of the disordered α-phase ($\eta_\alpha$=0), $W$ is an energy scale parameter introduced for convenience. The equilibrium phases $\eta_\alpha$=0 and ordered (β-phase) $\eta_\beta$=1 are obtained as solutions, $\eta_e = 0,1$, of the MF equation of state $\partial f(\eta,\Delta)/\partial\eta = 0$, where the non-equilibrium free energy density $f(\eta) = 2(H-H_\alpha)/W$, reduced by a factor $W/2$, is produced by the uniform order parameter $\eta \leq 1$; see Fig. 1, where the quantity $f(\eta)$ is depicted for two characteristic values of the parameter Δ: zero and 0.5. The standard MF analysis shows a third solution of the MF equation of state, $\eta_{\max} = (1-\Delta)/2$, which corresponds to a maximum $f_{\max} = (1-\Delta)^3(\Delta+3)/48$ of the function $f(\eta)$ for any $\Delta^2 < 1$. The solution $\eta_\beta = 1$ corresponds to a global minimum $f_\beta = -\Delta/3$ for any $\Delta < 1$ and, therefore, describes a stable phase. For $\Delta < 1$, the solution $\eta_\alpha = 0$ corresponds to a relative minimum $f_\alpha = 0$ of $f(\eta)$ and describes a metastable phase. At $\Delta = 1$ both $\eta_\alpha$ and $\eta_{\max}$ are equal to zero and describe a double inflection point of the reduced density $f(\eta)$. In a close vicinity of the spinodal point, that is, $\eta_\alpha = 0, \Delta \to 1$, the fluctuations of the disordered α-phase are large due to flatness of the Hamiltonian density (see Fig. 1). For $\Delta = 0$ both minima of $f(\eta)$ are equal ($f_\alpha = f_\beta = 0$) and therefore, at $\Delta = 0$ the phases α and β coexist in equilibrium. This means that $\Delta = 0$ is the equilibrium point of the first order phase transition [3]. The most part of these features of the model (1)-(2) are seen in Fig. 1. The relation between the MF free energy $F = VH(\eta_e, \Delta)$ and the reduced energy density $f(\eta) = 2(H - H_\alpha)/W$ can easily be obtained: $F = V[H_\alpha + Wf(\eta)/2]$. Having in mind that $f_\alpha = 0$ and $f_\beta = -\Delta/3$ we find the MF free energies of the phases $\alpha$ and $\beta$: $F_\alpha = VH_\alpha$, and $F_\beta = VH_\alpha - VW\Delta/6$.

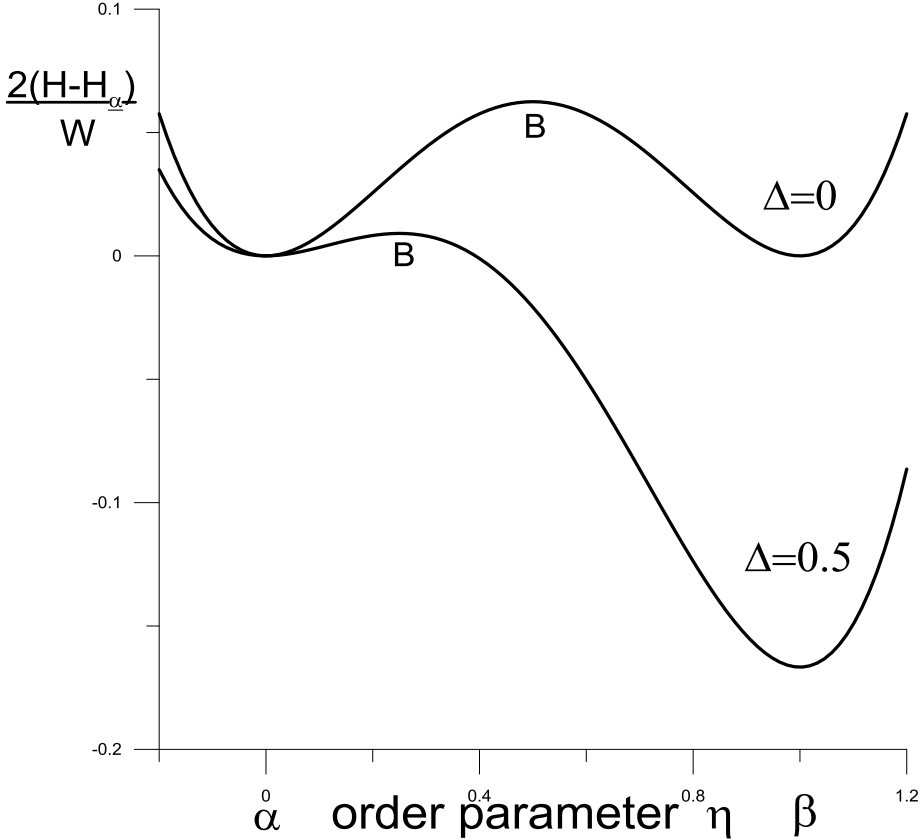

Figure 1. The Hamiltonian density in MFA, see Eq. (2), as a function of the uniform order parameter η for two values of the supersaturation Δ: zero and 0.5.

## 3. Fluctuation effects

Due to fluctuation effects the basic features of the systems described by the MF variant of model (1) - (2) undergo several modifications. Our consideration is restricted to the study of the fluctuation contributions to the order parameter, the internal and free energies. We study fluctuation effects in the disordered (α-) phase in the domain of metastability $0 < \Delta < 1$ of the disordered (α-) phase. For this aim we set $\eta_e = 0$ and consider purely fluctuation Hamiltonian; hereafter $\eta(\mathbf{x}) \equiv \delta\,\eta(\mathbf{x})$. Firstly, we outline the perturbation theory for the GL functional (1)-(2). Secondly, we present the results in GA when the fluctuation interactions are neglected. Thirdly, we outline the basic stages of the calculations with the help of the perturbation expansion up to the second order in of the interaction part of the GL Hamiltonian and present the final results of the perturbation theory.

### 3.1 General scheme of calculations

For our purposes we present the Hamiltonian (1)-(2) in the form

$$\mathcal{H}[\eta(\mathbf{x})] = H_\alpha V + \mathcal{H}_0[\eta(\mathbf{x})] + \mathcal{H}_{\text{int}}[\eta(\mathbf{x})], \qquad (3)$$

where $H_\alpha V$ is the α-phase free energy,

$$\mathcal{H}_0[\eta(\mathbf{x})] = \frac{1}{2}\int_V d^3x \,\{\kappa[\nabla\eta(\mathbf{x})]^2 + W(1-\Delta)\eta^2(\mathbf{x})\} \qquad (4a)$$

is the quadratic ("Gaussian") part describing free fluctuations, and $\mathcal{H}_{int} = (\mathcal{H}_3 + \mathcal{H}_4)$ - the interaction part, is given by two terms:

$$\mathcal{H}_3[\eta(\mathbf{x})] = \frac{\Delta-3}{3} W \int_V d^3x \, \eta^3(\mathbf{x}) \qquad (4b)$$

and

$$\mathcal{H}_4[\eta(\mathbf{x})] = \frac{1}{2} W \int_V d^3x \, \eta^4(\mathbf{x}). \qquad (4c)$$

The statistical average $\langle Q[\eta(\mathbf{x})]\rangle$ of any physical quantity $Q[\eta(\mathbf{x})]$ is defined by

$$\langle Q[\eta(\mathbf{x})]\rangle = \frac{1}{Z}\int D\eta Q[\eta(\mathbf{x})] \, e^{-\beta\mathcal{H}[\eta(\mathbf{x})]}, \qquad (5)$$

where $\beta = 1/k_B T$, $k_B$ is Boltzmann's constant, $T$ is the temperature, and

$$Z(\Delta, \beta) = \int D\eta e^{-\beta\mathcal{H}[\eta(\mathbf{x})]} \qquad (6)$$

is the partition function. In Eqs. (5) and (6), $\int D\eta$ denotes the functional integration $\prod_{\mathbf{x}\in V}\int d\eta(x)$ over all possible configurations of the field $\eta(\mathbf{x})$. For short, in the reminder of this paper we will often denote various functionals, for example, $Q[\eta(\mathbf{x})]$ and $\mathcal{H}[\eta(\mathbf{x})]$, by omitting the argument **x** of the field $\eta(\mathbf{x})$, $Q[\eta]$ and $\mathcal{H}[\eta]$, or, even by omitting the argument at all: $Q$ and $\mathcal{H}$.

The functional integrals in Eqs. (5) and (6) cannot be exactly calculated unless we apply the so-called Gaussian approximation (GA): $\mathcal{H}\{\eta\} \approx \mathcal{H}_0\{\eta\}$; the fluctuation interaction term $\mathcal{H}_{int}$ is ignored. To calculate the statistical average in GA, $\langle Q \rangle_0$, of any quantity $Q\{\eta\}$, or the partition function $\mathcal{Z}(\Delta, \beta)$ we should substitute $\mathcal{H}\{\eta\}$ in Eqs. (5) and (6) by $\mathcal{H}_0\{\eta\}$. The respective Gaussian partition function, $\mathcal{Z}_0(\Delta, \beta)$, can be exactly calculated by the Gaussian integral

$$\mathcal{Z}_0 = \int D\eta \, e^{-\beta \mathcal{H}_0[\eta]}. \tag{7}$$

GA is unreliable when the fluctuation interactions are relatively strong. This usually happens near critical points (continuous phase transitions) and spinodal points of first order phase transitions [3]. In our case, this occurs in a close vicinity of the spinodal point ($\Delta \sim 1$). In order to perform a more thorough study of the fluctuation effects we should take into account the fluctuation interactions by developing a perturbation expansion for the relevant physical quantities [2, 3].

The perturbation expansion is developed on the basis of statistical ensemble of free (non-interacting) fluctuations, namely, the ensemble defined by $\mathcal{H}_0\{\eta\}$. For example, the exponential statistical weigh in Eqs. (5) and (6) is expanded in infinite series in powers of the interaction part $\mathcal{H}_{int}$, or, more precisely, in powers of the dimensionless energy $\beta \mathcal{H}_{int}$. This perturbation expansion is performed also for the total partition function $\mathcal{Z}$, the free energy $\mathcal{F} \equiv -\beta^{-1} \ln \mathcal{Z}$, and any other quantities of interest. In practical calculations the perturbation series is truncated at some well justified order of $\beta \mathcal{H}_{int}$. In our study, the most essential results can be obtained by a truncation of the perturbation series just after the second power of $\beta \mathcal{H}_{int}$. While we do not intend investigation of the close vicinity of the spinodal point $\Delta = 1$, where the higher order terms might be relevant, this second order approximation for the perturbation series is enough for our purposes. In the remainder of this section we outline the main stages of the perturbation calculation and present the final results.

Firstly, we calculate the average of interest, for example, $\langle Q[\eta] \rangle$, in the exactly soluble GA of non-interacting order parameter fluctuations ($\mathcal{H}_{int} = 0$), namely, the first step is the calculation of $\langle Q[\eta] \rangle_o$. Secondly, we expand the exponents in Eq. (5):

$$\langle Q \rangle = \frac{\langle Q \, e^{-\beta \mathcal{H}_{int}} \rangle_0}{\langle e^{-\beta \mathcal{H}_{int}} \rangle_0} = \langle\langle Q \rangle\rangle - \beta \langle\langle Q \mathcal{H}_{int} \rangle\rangle + \frac{\beta^2}{2} \langle\langle Q \, \mathcal{H}_{int}^2 \rangle\rangle + \ldots \tag{8}$$

We have represented the r.h.s. of Eq. (8) by the so-called connected Gaussian averages $\langle\langle \ldots \rangle\rangle$. These averages are interrelated with the usual Gaussian averages $\langle \ldots \rangle_0$ [2, 3]. Some interrelationships of lowest order are given by

$$\langle\langle Q\rangle\rangle \equiv \langle Q\rangle_0 = \frac{1}{Z_0}\int D\eta\, Q\, e^{-\beta\mathcal{H}_0}, \qquad (9a)$$

$$\langle\langle QR\rangle\rangle = \langle QR\rangle_0 - \langle Q\rangle_0\langle R\rangle_0 \qquad (9b)$$

and

$$\langle\langle QRP\rangle\rangle \equiv \langle QRP\rangle_0 - \langle QR\rangle_0\langle P\rangle_0 - \langle QP\rangle_0\langle R\rangle_0 - \langle RP\rangle_0\langle Q\rangle_0$$

$$+ 2\langle Q\rangle_0\langle R\rangle_0\langle P\rangle_0\langle R\rangle, \qquad (9c)$$

where $Q[\eta(\mathbf{x})]$, $R[\eta(\mathbf{x})]$ and $P[\eta(\mathbf{x})]$ are monomials of the field $\eta(\mathbf{x})$. Note that within the basic ensemble defined by the quadratic Hamiltonian part $\mathcal{H}_0[\eta]$, which is invariant under the transformation $\eta(\mathbf{x}) \to -\eta(\mathbf{x})$, the Gaussian average of any odd monomial of $\eta(\mathbf{x})$ is equal to zero: $\langle\eta^{2n-1}(\mathbf{x})\rangle_0 = 0$, $(n > 0)$.

Furthermore, we present the total partition function of the system as a triple product $Z = Z_\alpha Z_0 Z_{int}$, where $Z_\alpha = \exp(-\beta H_\alpha V)$, $Z_0$ is given by Eq. (7), and

$$Z_{int} = \langle e^{-\beta\mathcal{H}_{int}\{\eta\}}\rangle_0 \qquad (10)$$

is entirely interaction product.

### 3.2 Perturbation expansion of several physical quantities

Within this scheme we calculate the ensemble averages, the overall (volume-averaged) order parameter, the internal and free energies. The volume-averaged order parameter is defined by

$$\eta_V = \frac{1}{V}\int_V d^3x\, \eta(\mathbf{x}). \qquad (11)$$

The Gaussian average of this quantity is equal to zero ($\langle\eta_V\rangle_0 = 0$) because of the first degree of the field $\eta(\mathbf{x})$ in Eq. (11). Hence, the mean value $\langle\eta_V\rangle$ of the overall order parameter $\eta_V$ is totally due to mode interactions,

$$\langle\eta_V\rangle \approx -\beta\langle\langle\eta_V\mathcal{H}_{int}\rangle\rangle + \frac{\beta^2}{2}\langle\langle\eta_V\mathcal{H}_{int}^2\rangle\rangle + \ldots, \qquad (12)$$

in particular, to the interaction of type $\eta^3(\mathbf{x})$; see Eq. (4b).

The perturbation expansion of the internal energy $E \equiv \langle\mathcal{H}\rangle = -\partial\ln Z/\partial\beta$ is given by

$$\mathcal{E} = \frac{1}{Z}\int D\eta\, \mathcal{H} e^{-\beta \mathcal{H}\{\eta\}} = \mathcal{E}_\alpha + \mathcal{E}_0 + \mathcal{E}_{int}, \tag{13}$$

where $E_\alpha = F_\alpha = VH_\alpha$ is the internal energy of phase α. This energy is equal to the free energy $F_\alpha$, calculated in Sec. 2. The fluctuation contribution to the internal energy in GA can be written in two alternative forms, namely,

$$\mathcal{E}_0 = -\frac{\partial}{\partial \beta}\ln Z_0 = \langle \mathcal{H}_0 \rangle_0, \tag{14}$$

and, therefore, could be calculated either with the help of Eq. (7) for $Z_0$, or, alternatively, as Gaussian average of the quadratic part $\mathcal{H}_0$ of the Hamiltonian. The fluctuation-interaction contribution up to the second order in $\beta \mathcal{H}_{int}$ is given by the truncated series

$$\mathcal{E}_{int} \approx \langle \mathcal{H}_{int}\rangle_0 - \beta \langle\langle \mathcal{H}_0 \mathcal{H}_{int}\rangle\rangle - \beta \langle\langle \mathcal{H}_{int}^2\rangle\rangle + \frac{\beta^2}{2}\langle\langle \mathcal{H}_0 \mathcal{H}_{int}^2\rangle\rangle. \tag{15}$$

The Helmholtz free energy $\mathcal{F} \equiv -\beta^{-1}(\ln Z)$ is given by the sum

$$\mathcal{F} = \mathcal{F}_\alpha + \mathcal{F}_0 + \mathcal{F}_{int}, \tag{16}$$

where $\mathcal{F}_\alpha = \mathcal{E}_\alpha = VH_\alpha$ is the free energy of phase α,

$$\mathcal{F}_0 = -\beta^{-1}\ln Z_0 \tag{17}$$

is the free energy in GA, and

$$\mathcal{F}_{int} = -\beta^{-1}\ln\langle e^{-\beta \mathcal{H}_{int}\{\eta\}}\rangle_0 \approx \langle \mathcal{H}_{int}\rangle_0 - \frac{\beta}{2}\langle\langle \mathcal{H}_{int}^2\rangle\rangle \tag{18}$$

is the fluctuation-interaction contribution to the free energy up to the second order of the perturbation expansion.

### 3.3 Calculations in the space of wave vectors

It is more convenient to calculate the averages in the reciprocal space of the wave vectors **k**. So, we apply the Fourier transformations of the order parameter field,

$$\eta(\mathbf{x}) = \sum_{\{\mathbf{k}\}} \eta(\mathbf{k}) e^{i\mathbf{k}\cdot\mathbf{x}}, \tag{19a}$$

$$\eta(\mathbf{k}) = \frac{1}{V} \int_V d^3 x\, \eta(\mathbf{x})\, e^{-i\mathbf{k}\cdot\mathbf{x}} \tag{19b}$$

to any quantity of interest. While we are interested only on bulk properties in a macroscopic cube of large volume $V = L^3$, we adopt periodic boundary conditions for the field $\eta(\mathbf{k})$ in wave vector ($\mathbf{k}$-) representation. Thus the wave vector $\mathbf{k} \equiv \{k_j, j = 1,2,3\}$ has components $k_j = 2\pi n_j / L ; n_j = 0, \pm 1,...$ The independent modes $\eta(\mathbf{k})$ are those with wave vectors $\mathbf{k}$ in the first Brillouin zone: $-\pi/a < k_j \leq \pi/a$ ($a > 0$ is the minimum length to which the study could be expanded). This means that the values of $n_j$ corresponding to independent Fourier amplitudes $\eta(\mathbf{k})$ are restricted to $n_j = 0, \pm 1,...,[L/a]$; $[L/a]$ is the maximal integer satisfying the requirement $[L/a] \leq L/a$. Therefore, the total number $N$ of independent fluctuation modes $\eta(\mathbf{k})$ in a large cube of size $L \gg a$ can be equalized to $(L/a)^3$:

$$N = \frac{V}{a^3} \gg 0. \tag{20}$$

This means that the number of independent modes $\eta(\mathbf{k})$ is equal to the maximal number of cubes of size $a$ which can be contained in the volume of the system. In microscopic investigations $a$ is the lattice constant (the mean inter-particle distance in amorphous bodies and fluids). In this case the number $N$ is identified with the number of particles in the body. But here the study is based on a quasi-macroscopic model and in order to keep the consistency of this study we should choose the minimum length scale under consideration with some caution, as we do at a next stage of our consideration.

The Hamiltonian can be expressed by the field amplitudes $\eta(\mathbf{k})$:

$$\mathcal{H}_0\{\hat{\eta}(\mathbf{k})\} = \frac{1}{2\beta} \sum_{\{\mathbf{k}\}} G_0^{-1}(\mathrm{k}) |\hat{\eta}(\mathbf{k})|^2, \tag{21a}$$

$$\mathcal{H}_3\{\hat{\eta}(\mathbf{k})\} = \vartheta \sum_{\{\mathbf{k}_1,\mathbf{k}_2\}} \hat{\eta}(\mathbf{k}_1)\hat{\eta}(\mathbf{k}_2)\hat{\eta}(-\mathbf{k}_1 - \mathbf{k}_2), \tag{21b}$$

$$\mathcal{H}_4\{\hat{\eta}(\mathbf{k})\} = u \sum_{\{\mathbf{k}_1,\mathbf{k}_2,\mathbf{k}_3\}} \hat{\eta}(\mathbf{k}_1)\hat{\eta}(\mathbf{k}_2)\hat{\eta}(\mathbf{k}_3)\hat{\eta}(-\mathbf{k}_1 - \mathbf{k}_2 - \mathbf{k}_3), \tag{21c}$$

where

$$G_0(\mathrm{k}) = \frac{R_0^3 \varepsilon}{V\delta(1+R_C^2 k^2)}, \tag{22}$$

$$\vartheta \equiv \frac{\Delta-3}{3} WV, \quad u \equiv \frac{1}{2} WV. \tag{23}$$

In Eqs. (21) - (23), $k = |\mathbf{k}|$, $\delta = 1 - \Delta$, $R_0 = (\kappa/W)^{1/2}$, $R_C = R_0/\delta^{1/2}$ is the correlation length (radius) of the fluctuations $\eta(\mathbf{k})$, and $\varepsilon = 1/\beta W R_0^3$ (in a slightly different form: $\varepsilon = T/k_B^{-1} W R_0^3$) is a rescaled temperature which participates in the perturbation expansion. Note that the quantity $G_0(k)$, given by Eq. (22), is the Fourier image of the pair correlation function in GA: $G_0(\mathbf{k}) = \langle |\eta(\mathbf{k})|^2 \rangle_o$ which depends on the magnitude $k = |\mathbf{k}|$ of the wave vector $\mathbf{k}$ (a result of the adopted isotropy of the systems under consideration).

A discussion of the important lengths $R_C$ and $R_0$ could be quite useful for our further analysis. For $\Delta = 0$, we have $R_C = R_0$ and, hence, the quantity $R_0$ could be interpreted as the minimal correlation radius of the fluctuations at the point ($\Delta=0$) of equilibrium ($\alpha - \beta$) phase transition (see Fig. 1). In the usual case of phase transitions driven by temperature variations the role of the factor $\delta^{-1/2}$ in $R_C$ is played by $|(T-T_C)/T_C|^{-1/2} > 0$, where $T_C$ is the equilibrium phase transition temperature. In this case, the correlation length $R_C(T)$ has a minimal size at $T=0$ and is called "zero-temparature correlation length", or, sometimes, coherence length: $R_C(0)$. This length defines a limit of application of the theory: The GL approach cannot be applied to studies of phenomena of characteristic size lower than the length $R_C(0)$. In our case this is the "coherence radius" $R_0$. Within our model, the length $R_0$ is therefore a candidate for a minimal length previously denoted by $a$.

From point of view of microphysics of particles or spin lattices, the length $R_0$ can be identified with the radius ($R_{int}$) of inter-particle or inter-spin interaction responsible for the phase transition. This is seen from derivations of GL functional from microscopic Hamiltonians (see, e.g., Ref. [3]). A length which is equal or lower than the radius of particle interaction $R_{int}$ is the mean inter-particle distance $a_0$, or, that is the same, the lattice constant of crystal bodies. For the most common case of short-range inter-particle interactions ($R_{int} \sim a_0$), we may think for the minimal length $R_0$ in our model as the radius of interaction of real particles in the system. Alternatively, if the system is divided in mesoscopic blocks (domains) of equal size $a$, we could interpret the length $R_0$ as the radius of short-range interactions between the blocks. A block may contain a single particle, several particles and even several tens or hundreds of particles and form a quasi-macroscopic system according to the standard macroscopic approach of description. The essential point is that the number of blocks should be enough large to ensure a justification of the statistical mechanics of fluctuations and an approximate but reliable application of the continuum limit ($L \gg a$). We assume that the number of blocks is $N = V/a^3$ is large, as given by Eq. (20) and the continuum limit could always be applied. Under these stipulations it seems reasonable to adopt $R_0$ for the interaction radius of blocks of size $a$ and assume, in case of short-range inter-block interaction, that $R_0 \sim a$ (almost "nearest neighbor block interactions"). In the extreme case of a block per particle, we have to use the relation $R_{int} \sim a_0$. Note, that expansion in the ratio $a/R_{int}$ has been considered in Ref. [7] and, more recently, in Ref. [8].

The functional integration $\int D\eta = \prod_{\mathbf{k}} \int d\eta(\mathbf{k})$ in integrals like those given by Eqs.(7) and (9a) can easily be performed. For this reason we should know precisely the full set of the independent complex amplitudes $\eta(\mathbf{k})$ and the real fields contained in such a set. As the field $\eta(\mathbf{x})$ is real, we have $\eta^*(\mathbf{k}) = \eta(-\mathbf{k})$; hence, the mode $\eta(\mathbf{0})$ is real and the modes $\eta(\mathbf{k})$ and $\eta(-\mathbf{k})$ with $\mathbf{k} \neq \mathbf{0}$ are not independent each other. The amplitudes $\eta(\mathbf{k})$ with $\mathbf{k} \neq \mathbf{0}$ can however be divided into two alternative sets of independent modes: those that belong to the upper half ($\mathbf{k} > \mathbf{0}$) of the Brillouin zone and those that pertain to the lower half ($\mathbf{k} < \mathbf{0}$). For definiteness, we chose the former. Furthermore, it is convenient to transform the complex modes $\eta(\mathbf{k})$ with $\mathbf{k} > \mathbf{0}$ to the real modes $a(\mathbf{k})$ and $b(\mathbf{k})$:

$$\eta(\mathbf{k}) = \frac{1}{\sqrt{2}}[a(\mathbf{k}) + ib(\mathbf{k})], \tag{24a}$$

and

$$\eta^*(\mathbf{k}) = \frac{1}{\sqrt{2}}[a(\mathbf{k}) - ib(\mathbf{k})]. \tag{24b}$$

Obviously, a full set of independent integration fields is identified with the following series of real field: $\eta(\mathbf{0})$, $a(\mathbf{k})$ and $b(\mathbf{k})$ with $\mathbf{k} > 0$. Within this choice of independent fields, the functional integration over the $\mathbf{k}$-modes should be performed according the rule

$$\int D\eta = \int_{-\infty}^{\infty} d\eta(\mathbf{0}) \prod_j \prod_{k_j > 0}^{k_j \leq \frac{\pi}{a}} \int_{-\infty}^{\infty} da(\mathbf{k}) \, db(\mathbf{k}). \tag{25}$$

Using the rule (25), the simple Gaussian integral in Eq. (7) can be easily calculated [3]. The result for the Gaussian partition function is given by

$$Z_0 = \prod_{\mathbf{k}}[2\pi G_0(k)]^{\frac{1}{2}}. \tag{26}$$

The most part of the further calculations are performed in the continuum limit $V \gg a$ in which the summation over the wave vector can be substituted by integration by the rule $\sum_{\mathbf{k}} \to V/(2\pi)^3 \int d^3k$. Besides, the calculation of the integrals is easier in spherical coordinates so we should define the upper cutoff $\Lambda$ for the magnitude $k$ of the wave vector $\mathbf{k}$: $k \leq \Lambda$. The cutoff $\Lambda$ should correspond to the number of blocks $N$, given by Eq. (20) and the sum $\sum_{\mathbf{k}} 1 = N$, where the wave vector $\mathbf{k}$ runs all points in the first Brillouin zone. Performing the continuum limit in this sum and solving the respective simple triple integral over $\mathbf{k}$ in Cartesian coordinates we immediately obtain the number $N$. Repeating the integration in spherical coordinates with cutoff $\Lambda$ and using the relation (20), we obtain the quantity $N(\Lambda a)^3/6\pi^2$. The latter should be equal to the number $N$. This requirement leads to the relation $(\Lambda a)^3/6\pi^2 = 1$ and, hence, to the appropriate cutoff $\Lambda = (6\pi^2)^{1/3}/a$. Now we are in the position to perform the study in the continuum limit for large systems ($V \gg a^3$) without any inconsistencies between the possible variants of calculation.

### 3.4. Results in Gaussian approximation

From Eqs. (14), (22) and (26) we obtain the internal energy in GA:

$$\mathcal{E}_0 = \frac{V}{2\beta a^3}, \tag{27}$$

or, alternatively, $E_0 = N/2\beta$. Note that this result is in accord with the equipartition theorem.

The free energy in GA follows from Eqs. (17) and (26):

$$\mathcal{F}_0 = -\frac{1}{2\beta}\sum_{\{\mathbf{k}\}} \ln[2\pi G_0(k)], \tag{28}$$

where $G_0(k)$ is given by Eq. (22). In the continuum limit, we have $F_0 = -A_0/2\beta$, where the integral $A_0$ is given by

$$A_0 = \frac{V}{(2\pi)^3}\int d^3k \ln[2\pi G_0(k)]. \tag{29}$$

Using Eq. (22) for $G_0(k)$ and calculating the integral $A_0$ with the cutoff $\Lambda = (6\pi^2)^{1/3}/a$, we obtain the free energy $\mathcal{F}_0$ in the form

$$\mathcal{F}_0 = \mathcal{E}_0\left[\ln\frac{V\delta(1+U^2)}{2\pi R_0^3 \varepsilon} - \frac{2}{3} + \frac{2}{U^2} - 2\frac{\arctan U}{U^3}\right], \tag{30}$$

where

$$U = \Lambda R_C = \frac{(6\pi^2)^{1/3} R_0}{a\delta^{1/2}} \tag{31}$$

is the dimensionless cutoff. In a close vicinity of the spinodal point ($\delta \ll 1$), where $U \gg 1$, we have

$$\mathcal{F}_0 \approx \mathcal{E}_0 \left( \ln \frac{V}{a^2 R_0 \varepsilon} + \frac{1}{3} \ln \frac{9\pi}{2} - \frac{2}{3} \right). \tag{32}$$

### 3.5. Perturbation theory results

To calculate the interaction corrections to the physical quantities we apply the standard methods of the perturbation theory [2, 3]. Bearing in mind that the Gaussian averages of odd products of the field $\eta(\mathbf{k})$ are equal to zero, we can write the Eq. (12) in the form

$$\langle \eta_V \rangle = -\beta \langle\langle \eta_V \mathcal{H}_3 \rangle\rangle + \beta^2 \langle\langle \eta_V \mathcal{H}_3 \mathcal{H}_4 \rangle\rangle, \tag{33}$$

where terms of order $O[(\beta \mathcal{H}_{int})^3]$ have been neglected.

The calculation of the averages in Eq, (33) leads to the following result

$$\langle \eta_V \rangle = -3\beta \vartheta G(0) A_1 + 12\beta^2 u \vartheta G_0(0) [G_0(0) A_1^2 + 3 A_1 A_2 + 2 C_3], \tag{34}$$

where the perturbation integrals $A_1$ and $A_2$ are given by the general formula

$$A_m = \frac{V}{(2\pi)^3} \int d^3 k \, G_0^m(k), \tag{35}$$

($m = 1,2,...$), and the integral $C_3$ has the form

$$C_3 \equiv \frac{V^2}{(2\pi)^6} \int d^3k \, d^3p \, G_0(k) G_0(p) G_0(|\mathbf{k}+\mathbf{p}|). \tag{36}$$

In order to perform a correct study we have to understand the leading dependence of the perturbation terms on the parameters $\varepsilon$ and $\delta$. Apart of any calculations we could understand this with the help of the following qualitative analysis. The shape of the correlation function (22) implies that $G_0(k) \sim \varepsilon/\delta$ and $k^2 \sim \delta$. The last relation indicates that $k^2 dk \sim \delta^3$. Then one easily obtains that the leading term in $A_1$ is of type $A_1 \sim \varepsilon \delta^0 \sim \varepsilon$. For the other two integrals in Eq. (34) we obtain $A_2 \sim \varepsilon^2 / \delta^{1/2}$ and $C_3 \sim \varepsilon^3 \delta^0$ ($\delta^0$ denotes either independence or $log$-dependence on the parameter δ). Further, bearing in mind that Eq. (23) for the parameters $u$ and $\vartheta$ together with the relation $\varepsilon = 1/\beta W R_0^3$ yield $u \sim \vartheta \sim \varepsilon^{-1}$, we easily find that the first-order perturbation term in Eq. (34) is proportional to $\varepsilon/\delta$ whereas the first, the second and the third $u\vartheta$ - terms in the same equation are proportional to $\varepsilon^2/\delta^2$, $\varepsilon^2/\delta^{3/2}$ and $\varepsilon^2\delta^0$, respectively. For $\varepsilon/\delta \ll 1$ the contribution from the second-order perturbation terms is smaller than the first-order contribution, and our perturbation theory is well definite.

In fact, this theory has two effective expansion parameters, $\varepsilon/\delta$ and $\varepsilon/\delta^{1/2}$, which correspond to the interaction terms $\mathcal{H}_3$ and $\mathcal{H}_4$, respectively. For $\delta < 1$, $\varepsilon/\delta > \varepsilon/\delta^{1/2}$ and, hence, the ratio $\varepsilon/\delta$ is the major expansion parameter. The latter is produced through the $\eta^3$-interaction, whereas the effective expansion parameter $\varepsilon/\delta^{1/2}$ is due to the presence of the $\eta^4$-interaction. The critical region, where the perturbation terms become large and the theory breaks down, can approximately be defined by $\varepsilon > \delta$. Here we are interested on fluctuation properties outside this region, where $\varepsilon \ll \delta$. Thus we consider almost Gaussian fluctuations when the fluctuation interactions are weak and their effect can be taken into account by the lowest order perturbation terms. For $\varepsilon \ll \delta$, we may assume the parameter ε as expansion parameter within our study.

Another important feature of this quite complex theory is the following. At first glance one may suppose that Eq. (34) contains all terms of orders $\varepsilon^1$ and $\varepsilon^2$. While this is true for the $\varepsilon^1$-order, other two $\varepsilon^2$-term of type $\varepsilon^2\delta^{-5/2}$ and $\varepsilon^2\delta^0$ are generated from the third order perturbations, $\beta^3 \langle\langle \eta_V \mathcal{H}_{int}^3 \rangle\rangle$, more specifically, from $\langle\langle \eta_V \mathcal{H}_3^3 \rangle\rangle$. Thus the $m$-th order perturbation term $\beta^m \langle\langle \eta_V \mathcal{H}_{int}^m \rangle\rangle$ with $m > 1$ does not contain all terms of order $\varepsilon^m$. The reason is in the different number of fields in the interaction terms. Consider the average $\langle\langle \eta_V \mathcal{H}_4^n \mathcal{H}_3^l \rangle\rangle \neq 0$ of order $m = n + l > 1$, where $n \geq 0$ and $l$ is odd integer. Certainly, this average generates $G_0$-functions of number $(4n + 3l + 1)/2$. Having in mind that the interaction parameters $u$ and $\vartheta$ are of order $\varepsilon^{-1}$ we obtain that the respective average generates $(\varepsilon^\rho/\delta^\zeta)$-terms with exponent $\rho = n + (l+1)/2$. Thus the order $m$ of the perturbation term coincides ($m = \rho$) with the order $\rho$ of the expansion parameter ε, only if $l = 1$. The value of the exponent $\zeta$ depends on the degree of connection between the multipliers in the average $\langle\langle \eta_V \mathcal{H}_4^n \mathcal{H}_3^l \rangle\rangle$. Using the established rules, we may check that the first term in Eq. (34) produces exponent $\rho = 1$ and the three second order terms generate exponent $\rho = 2$. This is in conformity with Eq. (33), where the first term corresponds to $n = 0$ and $l = 1$, whereas the second term is characterized by $n = l = 1$.

Using the same arguments, one may easily evaluate the $m$-th order perturbation terms of type $\langle\langle \eta^{2\sigma} \mathcal{H}_4^n \mathcal{H}_3^{2\nu} \rangle\rangle \neq 0$, where $(n + 2\nu) = m$, $\sigma$ and $\nu$ are nonnegative integer numbers. We easily obtain that this average generates $(\varepsilon^{\hat{\rho}}/\delta^{\hat{\zeta}})$- terms, where $\hat{\rho} = (\sigma + n + \nu)$ and value of the exponent $\hat{\zeta}$ depends on the way of connection of the three multiplier in the average. This type of averages are used in the calculation of the internal energy ($\sigma = 1$) and the free energy ($\sigma = 0$).

In the reminder of this paper we shall assume $\varepsilon \ll \delta$ and take into account only the perturbation terms of lowest order in $\varepsilon$. In this approximation, Eq. (34) takes the form

$$\langle \eta_V \rangle = -3\beta\vartheta G(0)A_1 + O(\varepsilon^2) \tag{37}$$

The solution of the integral $A_1$ is given by

$$A_1 = \frac{1}{\pi}\left(\frac{3}{4\pi}\right)^{1/3}\frac{R_0}{a}a_1(\delta)\varepsilon, \tag{38}$$

where the function $a_1(\delta)$, defined through

$$a_1[U(\delta)] = 1 - \frac{\arctan U(\delta)}{U(\delta)} \tag{39}$$

and Eq. (31), takes values $a_1 \leq 1$ for $\delta \in [0,1]$ and can be accepted as a number coefficient of order unity. Thus Eq. (37) can be written in the form

$$\langle \eta_V \rangle = \frac{a_1}{\pi}\left(\frac{3}{4\pi}\right)^{\frac{1}{3}}\frac{R_0}{a}\left(\frac{2}{\delta} + 1\right)\varepsilon + O(\varepsilon^2). \tag{40}$$

Following the same approach, we investigate the perturbation contributions to the free and internal energies. The calculation of the averages in Eq. (18) yields the interaction part of the free energy in the form

$$\mathcal{F}_{int} = 3uA_1^2 - \frac{9}{2}\beta\vartheta^2 G_0(0)A_1^2 - 3\beta\vartheta^2 C_3 - 36\beta u^2 A_1^2 A_2 - 12\beta u^2 C_4, \tag{41}$$

where

$$C_4 = \frac{V^3}{(2\pi)^9}\int d^3k\, d^3p\, d^3q\, G_0(k)G_0(p)G_0(q)G_0(|\mathbf{k}+\mathbf{p}+\mathbf{q}|). \tag{42}$$

The first three terms in the r.h.s. of Eq. (41) is of order $\varepsilon^1$, whereas the last two terms are of order $\varepsilon^2$; note that the integral (42) contains four correlation functions $G_0$ and, hence, is of order $\varepsilon^4$. The next orders of the perturbation theory contain terms of order $O(\varepsilon^2)$. Here we ignore all terns of order $\varepsilon^2$. After some algebra with the first three terms in Eq. (41), we obtain the result for the interaction part $\mathcal{F}_{int}$:

$$\mathcal{F}_{int} = \frac{3}{\pi^2}\left(\frac{3}{4\pi}\right)^{\frac{2}{3}}\left(\frac{a}{R_0}\right)a_1^2\varepsilon E_0\left[1 - \frac{(2+\delta)^2}{3\delta} - \frac{2\pi^2}{9}\left(\frac{4\pi}{3}\right)^{\frac{2}{3}}(2+\delta)^2\left(\frac{a}{R_0}\right)^2\frac{c_3}{a_1^2}\right], \tag{43}$$

where the dimensionless integral

$$C_3 = \int \frac{d^3x d^3y}{(2\pi)^6} \frac{1}{(1+x^2)(1+y^2)(1+|\mathbf{x}+\mathbf{y}|^2)} \tag{44}$$

is related with the integral (36) by the equality $C_3 = R_0^3 \varepsilon^3 c_3 / V$. In Eq. (44), $\mathbf{x} = R_C \mathbf{k}$ and $\mathbf{y} = R_C \mathbf{k}$ are dimensionless wave vectors with magnitudes $x \leq U$ and $y \leq U$.

The interaction part $E_{int}$ of the internal energy can be calculated from Eq. (15). Alternatively, this energy can be obtained with the help of the thermodynamic relation $E = d(\beta F)/d\beta$. The result is

$$E_{int} = -3uA_1^2 + \frac{9}{2}\beta\vartheta^2 G_0(0)A_1^2 + 3\beta\vartheta^2 C_3 + 72\beta u^2 A_1^2 A_2 + 24\beta u^2 C_4. \tag{45}$$

The terms in Eq. (45) are similar to those in Eq. (41) for the free energy $F_{int}$. The first three terms have opposite signs towards the respective terms in Eq. (41) whereas the last two terms have opposite signs and twice bigger number coefficients. Therefore, to first order in $\varepsilon$ we have $E_{int} = -F_{int}$, where $F_{int}$ is given by Eq. (43). The total internal energy to first order in $\varepsilon$ can therefore be written in the form

$$E = E_\alpha$$

$$+E_0 \left\{ 1 - \frac{3}{\pi^2} \left(\frac{3}{4\pi}\right)^{\frac{2}{3}} \left(\frac{a}{R_0}\right) a_1^2 \varepsilon \left[ 1 - \frac{(2+\delta)^2}{3\delta} - \frac{2\pi^2}{9} \left(\frac{4\pi}{3}\right)^{\frac{2}{3}} (2+\delta)^2 \left(\frac{a}{R_0}\right)^2 \frac{c_3}{a_1^2} \right] \right\},$$

$$\tag{46}$$

where $E_0$ is given by Eq. (27); see also Eq. (13).

## 4. Concluding remarks

We have calculated fluctuation corrections to the internal energy, the free energy and the overall order parameter in metastable states of first order phase transitions described by a third order term in the GL Hamiltonian. The Gaussian approximation leads to results typical for non-interacting particles [1]. The leading effect of the fluctuation interaction comes from the third order term in the GL Hamiltonian and the typical for this term expansion parameter $\varepsilon/\delta$. This is in a contrast with the usual case of second order phase transitions where the fluctuation interactions produce contributions in powers of $\varepsilon/\delta^{1/2}$. Our study includes the first non-vanishing contributions from the fluctuation interactions and is valid for $\varepsilon/\delta < 1$, i.e., outside a pseudo-critical region around the spinodal point $\delta = 0$, where the higher order perturbation contributions are large and the perturbation theory breaks down. It seems interesting to compare the present results with results from a suitable numerical analysis.